\documentclass{article}

\usepackage[preprint, nonatbib]{nips_2018}

\usepackage[utf8]{inputenc} 
\usepackage[T1]{fontenc}    
\usepackage{hyperref}       
\usepackage{url}            
\usepackage{booktabs}       
\usepackage{amsfonts}       
\usepackage{nicefrac}       
\usepackage{microtype}      
\usepackage{xcolor}
\usepackage{bm}
\usepackage{amsmath}
\usepackage{graphicx}
\graphicspath{{graph/}}
\usepackage{amsmath}
\usepackage{float}
\usepackage[export]{adjustbox}
\usepackage{subcaption}
\usepackage{makecell}

\usepackage{todonotes}

\title{Deep Learning for Portfolio Optimization}

\author{
  Zihao Zhang, Stefan Zohren, Stephen Roberts\\
  Oxford-Man Institute of Quantitative Finance,\\
  University of Oxford
}

\begin{document}

\maketitle

\begin{abstract}
We adopt deep learning models to directly optimise the portfolio Sharpe ratio. The framework we present circumvents the requirements for forecasting expected returns and allows us to directly optimise portfolio weights by updating model parameters. Instead of selecting individual assets, we trade Exchange-Traded Funds (ETFs) of market indices to form a portfolio. Indices of different asset classes show robust correlations and trading them substantially reduces the spectrum of available assets to choose from. We compare our method with a wide range of algorithms with results showing that our model obtains the best performance over the testing period, from 2011 to the end of April 2020, including the financial instabilities of the first quarter of 2020.  A sensitivity analysis is included to understand the relevance of input features and we further study the performance of our approach under different cost rates and different risk levels via volatility scaling.
\end{abstract}


\section{Introduction}
\label{introduction}

Portfolio optimisation is an essential component of a trading system. The optimisation aims to select the best asset distribution within a portfolio in order to maximise returns at a given risk level. This theory was pioneered in Markowitz's key work \cite{markowitz1952portfolio} and is widely known as modern portfolio theory (MPT).  The main benefit of constructing such a portfolio comes from the promotion of diversification that smoothes out the equity curve, leading to a higher return per risk than trading an individual asset. This observation has been proven (see e.g. \cite{zivot2017introduction}) showing that the risk (volatility) of a long-only portfolio is always lower than that of an individual asset, for a given expected return, as long as assets are not perfectly correlated. We note that this is a natural consequence of Jensen's inequality \cite{jensen1906fonctions}.

Despite the undeniable power of such diversification, it is not straightforward to select the ``right'' asset allocations in a portfolio, as the dynamics of financial markets change significantly over time. Assets that exhibit, for example, strong negative correlations in the past could be positively correlated in the future. This adds extra risk to the portfolio and degrades subsequent performance. Further, the universe of available assets for constructing a portfolio is enormous. Taking the US stock markets as a single example,  more than 5000 stocks are available to choose from \cite{wild2008index}. Indeed, a well rounded portfolio not only consists of stocks, but also is typically supplemented with bonds and commodities, further expanding the spectrum of choices.

In this work, we consider directly optimising a portfolio, utilising deep learning models \cite{lecun2015deep, goodfellow2016deep}. Unlike classical methods \cite{markowitz1952portfolio} where expected returns are first predicted (typically through econometric models), we bypass this forecasting step to directly obtain asset allocations. Several works \cite{moody1998performance, moody2001learning,  zhang2020deep} have shown that the return forecasting approach is not guaranteed to maximise the performance of a portfolio, as the prediction steps attempt to minimise a prediction loss which is not the overall reward from the portfolio. In contrast, our approach is to directly optimise the Sharpe ratio  \cite{sharpe1994sharpe}, thus maximising return per unit of risk. Our framework starts by concatenating multiple features from different assets to form a single observation and then uses a neural network to extract salient information and output portfolio weights so as to maximise the Sharpe ratio.

\begin{figure}[t]
\centering
\includegraphics[width=5.5in, height=2in]{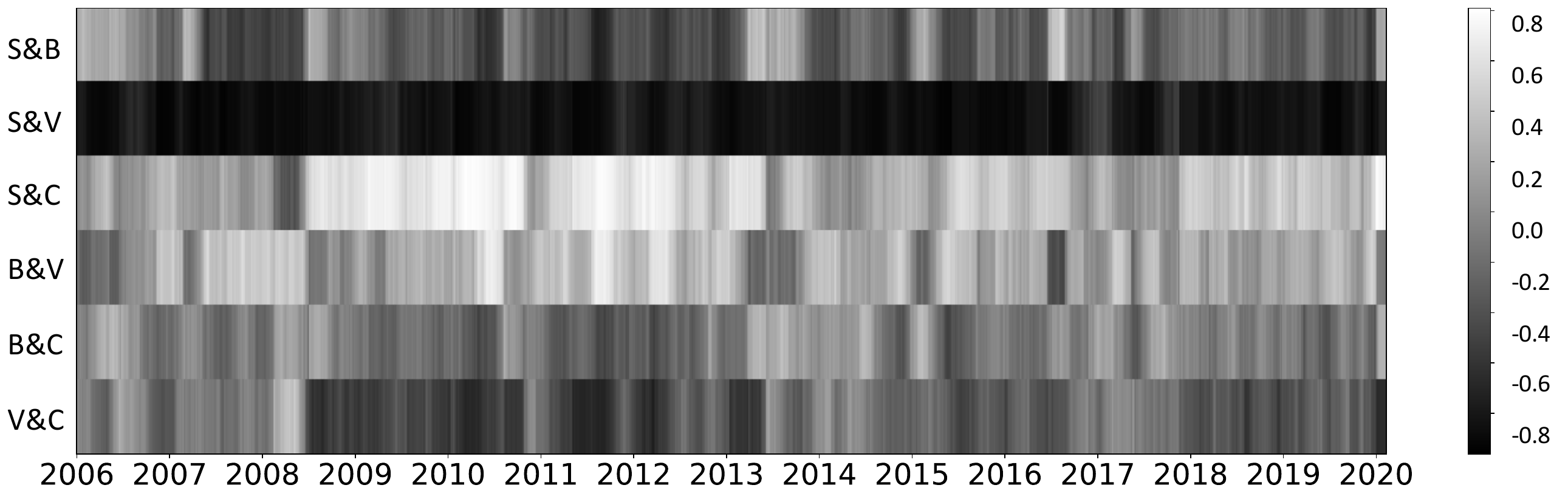}
\caption{Heatmap for rolling correlations between different index pair. (S: stock index, B: bond index, C: commodity index and V: volatility index.)}
\label{fig:correlation}
\end{figure}

Instead of choosing individual assets, Exchange-Traded Funds (ETFs) \cite{gastineau2008exchange} of market indices are selected to form a portfolio. We use four market indices: US total stock index (VTI), US aggregate bond index (AGG), US commodity index (DBC) and Volatility Index (VIX). All of these indices are popularly traded ETFs that offer high liquidity and relatively small expense ratios. Trading indices substantially reduces the possible universe of asset choices and gains exposure to most securities. Further, these indices are generally uncorrelated, or even negatively correlated, as shown in Figure~\ref{fig:correlation}. Individual instruments in the same asset class, however, often exhibit strong positive correlations. For example, more than 75\% stocks are highly correlated with the market index \cite{wild2008index}, thereby adding them to a portfolio helps less with diversification. 

We are aware that subsector indices can be included in a portfolio, rather than using the total market index, since sub-industries perform at different levels and a weighting on good performance in a sector would therefore deliver extra returns. However, we see subsector indices as highly correlated, thus adding them again provides minimal diversification for the portfolio, and risks lowering returns per unit risk. If higher returns are desired, we can use (e.g.) volatility scaling to upweight our positions and amplify returns. We therefore do not believe there is a need to find the best performing sector. Instead, we aim to provide a portfolio that delivers high return per unit risk, and allows for volatility scaling \cite{moskowitz2012time, harvey2018impact, lim2019enhancing} to achieve desired return levels.

\paragraph{Outline:}
The remainder of the paper is structured as follows. We introduce relevant literature in Section~\ref{literature} and present our methodology in Section~\ref{method}. Section~\ref{experiment} describes our experiments and details the results of our method compared with a range of baseline algorithms. In Section~\ref{conclusion}, we summarise our findings and discuss possible future work.

\section{Literature Review}
\label{literature}

In this section, we review popular portfolio optimisation methods and discuss how deep learning models have been applied to this field. There is a vast literature available on this topic, so we aim merely to highlight key concepts, popular in the industry or in academic study. One of the popular practical approaches is the reallocation strategy \cite{wild2008index} adopted by many pension funds (for example, LifeStrategy Equity Fund, Vanguard). This approach constructs a portfolio by only investing in stocks and bonds. A typical risk moderate portfolio would, for example, comprise 60\% equities and 40\% bonds and the portfolio needs to be only rebalanced semi-annually or annually to maintain this allocation ratio. The method delivers good performance over the long term, however the fixed allocation ratio means that investors with preference for more weight on stocks need to tolerate potentially large drawdowns during dull markets.

Mean-variance analysis or MPT \cite{markowitz1952portfolio} is used for many institutional portfolios that solves a constraint optimisation problem to derive portfolio weights. Despite its popularity, the assumptions of the theory are under criticism as they are often not obeyed in real financial markets. In particular, returns are assumed to follow a Gaussian distribution in MPT, therefore, investors only consider expected return and variance of the portfolio returns to make decisions.
However, it is widely accepted (see for instance \cite{cont1999statistical, zhang2019extending}) that returns tend to have fat tails and extreme losses are more likely to occur in practice, leading to severe drawdowns that are not bearable.
The Maximum Diversification (MD) portfolio is another promising method introduced in \cite{choueifaty2008toward} that aims to maximise the diversification of a portfolio, thereby aiming to have minimally correlated assets so the portfolio can achieve higher returns (and lower risk) than other classical methods. We compare our model with both these strategies, with the results suggesting that our methods deliver better performance and tolerate larger transaction costs than either of these benchmarks.


Stochastic Portfolio Theory (SPT) was recently proposed in \cite{fernholz2002stochastic, fernholz2009stochastic}. Unlike other methods, SPT aims to achieve relative arbitrages meaning to select portfolios that can outperform a market index with probability one. Such investment strategies have been studied in \cite{fernholz2010optimal, fernholz2011optimal, ruf2013hedging, wong2015optimization}. However, the number of relative arbitrage strategies remains small, as theory does not suggest how to construct such strategies. We can check whether a given strategy is a relative arbitrage, but it is non-trivial to develop one \textit{ex ante}. In this work, we 
include a particular class of SPT called functionally generated portfolio (FGP) \cite{fernholz1999portfolio} in our experiment, but the result suggests this method delivers inferior performance than other algorithms and generates large turnovers, making it unprofitable under heavy transaction costs. 

The idea of our end-to-end training framework was first initiated in \cite{moody1998performance, moody2001learning}. However, these works mainly focus on optimising the performance for a single asset so there is little discussion on how portfolios should be maximised. Furthermore, their testing period is from 1970 to 1994, whereas our dataset is up to date and we study the behavior of our strategy under the current crisis due to COVID-19. We can also link our approach to reinforcement learning (RL) \cite{sutton2018reinforcement, mnih2013playing, williams1992simple} where an agent interacts with an environment to maximise cumulative rewards. The works of \cite{bertoluzzo2012testing, huang2018financial, zhang2020deep} have studied this stream and adopted RL to design trading strategies. However, the goal of RL is to maximise expected cumulative rewards such as profits whereas Sharpe ratio can not be directly optimised.

\section{Methodology}
\label{method}

In this section, we introduce our framework and discuss how Sharpe ratio can be optimised through gradient ascent. We discuss the types of neural networks used and detail the functionality of each component in our method.  

\subsection{Objective Function}

The Sharpe ratio is used to gauge the return per risk of a portfolio and is defined as expected return over volatility (excluding risk-free rate for simplicity): 
\begin{equation}
L = \frac{E(R_{p})}{\mathrm{Std}(R_{p})} 
\end{equation}
where  $E(R_{p})$ and $\mathrm{Std}(R_{p})$ are the estimates of the mean and standard deviation of portfolio returns. Specifically, for a trading period of $t = \{1, \cdots, T \}$, we can maximise the following objective function:
\begin{equation}
\begin{split}
L_T &= \frac{E(R_{p,t})}{\sqrt{E(R^2_{p,t}) - (E(R_{p,t}))^2 }} \\
&E(R_{p,t}) = \frac{1}{T} \sum_{t=1}^T R_{p,t}
\end{split}
\end{equation}
where $R_{p,t}$ is realized portfolio return over $n$ assets at time $t$ denoted as:
\begin{equation}
R_{p,t} = \sum_{i=1}^n w_{i,t-1} \cdot r_{i,t}
\end{equation}
where $r_{i,t}$ is the return of asset $i$ with $r_{i,t}=(p_{i,t} / p_{i, t-1}-1)$. We represent the allocation ratio (position) of asset $i$ as $w_{i,t} \in [0,1] $ and $\sum_i^n w_{i,t} = 1$. In our approach, a neural network $f$ with parameters $\theta$ is adopted to model $w_{i,t}$ for a long-only portfolio:
\begin{equation}
w_{i,t} = f(\theta | x_t) 
\end{equation}
where $x_t$ represents the current market information and we bypass the classical forecasting step by linking the inputs with positions to maximise the Sharpe over trading period $T$, namely $L_T$. However, a long-only portfolio imposes constraints that require weights to be positive and summed to one, we use softmax outputs to fulfill these requirements:
\begin{equation}
w_{i,t} = \frac{\exp(\tilde{w}_{i,t})}{\sum_{j}^n \exp(\tilde{w}_{j,t})}, \quad \text{where} \ \tilde{w}_{i,t} \ \text{are the raw weights.}  
\end{equation}
Such a framework can be optimised using unconstrained optimisation methods. Particularly, we use gradient ascent to maximise the Sharpe ratio. The gradient of $L_T$ with respect to parameters $\theta$ is readily calculable, with an excellent derivation presented in \cite{moody1998performance, molina2016stock}. Once we obtain $\partial L_T / \partial \theta$, we can repeatedly compute this value from training data and update the parameters by using gradient ascent:
\begin{equation}
\theta_{new} :=\theta_{old} + \alpha \frac{\partial L_T}{\partial \theta}
\end{equation}
where $\alpha$ is the learning rate and the process can be repeated for many epochs until the convergence of Sharpe ratio or the optimisation of validation performance is achieved.

\subsection{Model Architecture}

\begin{figure}[t]
\centering
\includegraphics[width=3.5in, height=3.6in]{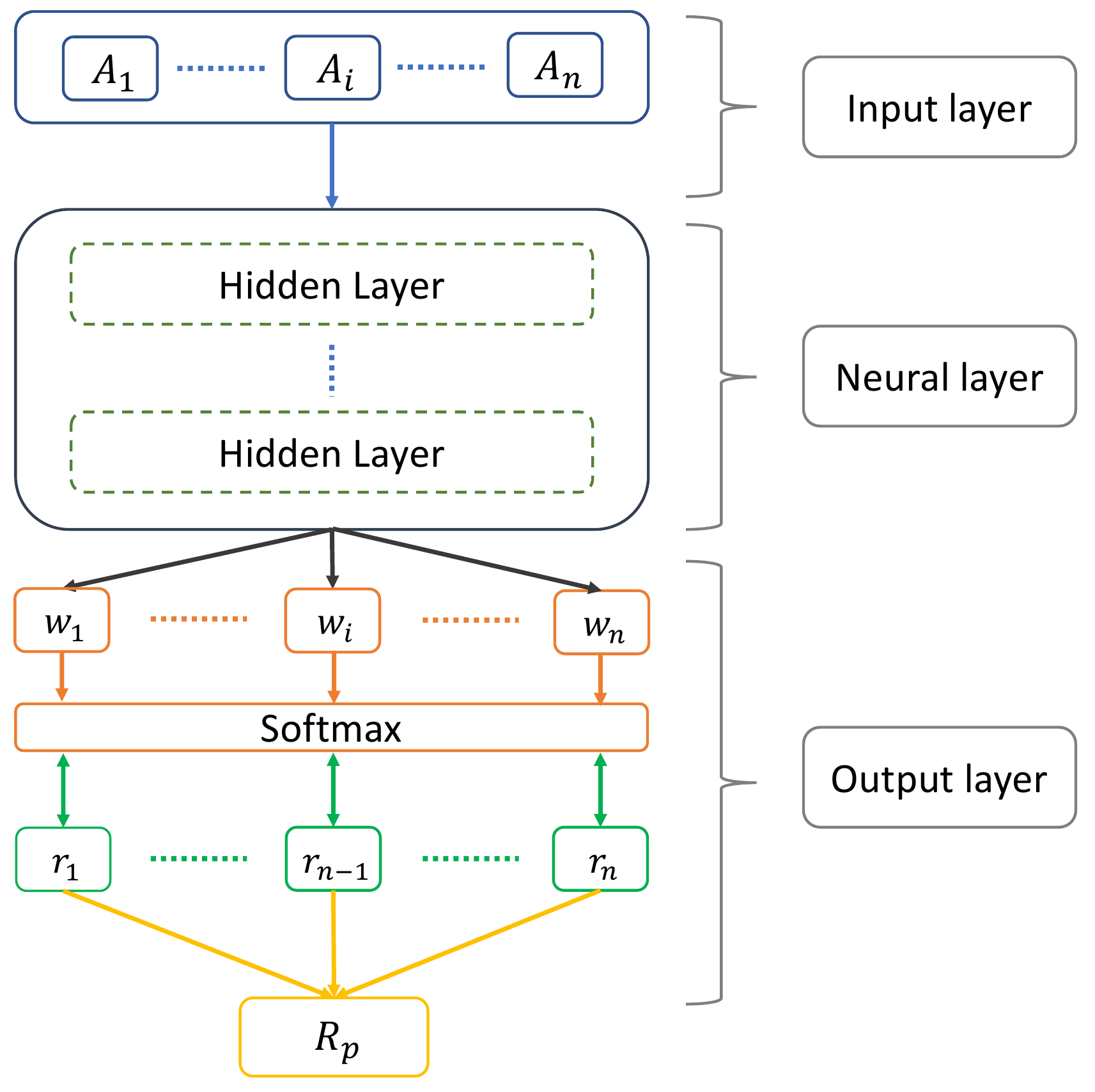}
\caption{Model architecture schematic. Overall, our model contains three main building blocks: input layer, neural layer and output layer.}
\label{fig:model}
\end{figure}

We depict our network architecture in Figure~\ref{fig:model}. Our model consists of three main building blocks: input layer, neural layer and output layer. The idea of this design is to use neural networks to extract cross-sectional features from input assets. Features extracted from deep learning models have been suggested to perform better than traditional hand-crafted features \cite{zhang2020deep}. Once features have been extracted, the model outputs portfolio weights and we obtain realised returns to maximise Sharpe ratio. The following details each component of our method. 

\paragraph{Input layer} 
We denote each asset as $A_i$ and we have $n$ assets to form a portfolio. A single input is prepared by concatenating information from all assets. For example, the input features of one asset can be its past prices and returns with a dimension of $(k, 2)$ where $k$ represents the lookback window. By stacking features across all assets, the dimension of the resulting input would be $(k, 2 \times n)$. We can then feed this input to the network and expect non-linear features being extracted. 

\paragraph{Neural layer}
A series of hidden layers can be stacked to form a network, however, in practice, this part requires lots of experiments as there are plentiful ways of combining hidden layers and the performance often depends on the design of architecture. We have tested deep learning models including fully connected neural network (FCN), convolutional neural network (CNN) and Long Short-Term Memory (LSTM) \cite{hochreiter1997long}. Overall, LSTMs deliver the best performance for modelling daily financial data and a number of works \cite{tsantekidis2017using, lim2019enhancing, zhang2020deep} support this observation. 

We note the problem of FCN is its problem of severe overfitting. As it assigns parameters to each input feature, this results in an excess number of parameters. The LSTM operates with a cell structure that has gate mechanisms to summarise and filter information from its long history, so the model ends up with fewer trainable parameters and achieves better generalisation results. In contrast, CNNs with a strong smoothing (typical of large convolutional filters) tend to have underfitting problems, such that oversmooth solutions are obtained. Due to the design of parameter sharing and the convolution operations, we experience CNNs to overfilter the inputs. However, we note that CNNs appear to be excellent candidates for modelling high-frequency financial data such as limit order books \cite{zhang2019deeplob}.

\paragraph{Ouput layer}
In order to construct a long-only portfolio, we use the \emph{softmax} activation function for the output layer, which naturally imposes constraints to keep portfolio weights positive and summing to one. The number of output nodes ($w_1, \cdots, w_n$) is equal to the number of assets in our portfolio, and we can multiply these portfolio weights with associated assets' returns ($r_1, \cdots, r_n$) to calculate realised portfolio returns $(R_{p})$. Once realised returns are obtained, we can derive the Sharpe ratio and calculate the gradients of the Sharpe ratio with respect to the model parameters and use gradient ascent to update the parameters. 

\section{Experiments}
\label{experiment}

\subsection{Description of Dataset}

We use four market indices: US total stock index (VTI), US aggregate bond index (AGG), US commodity index (DBC) and Volatility Index (VIX). These are popular Exchange-Traded Funds (ETFs) \cite{gastineau2008exchange} that have existed for more than 15 years. As discussed in Section~\ref{introduction}, trading indices offers advantages over trading individual assets because these indices are generally uncorrelated resulting in diversification. A diversified portfolio delivers a higher return per risk and the idea of our strategy is to have a system that delivers good reward-to-risk ratio. Our dataset ranges from 2006 to 2020 and contains daily observations. We retrain our model at every 2 years and use all data available up to that point to update parameters. Overall, our testing period is from 2011 to the end of April 2020, including the most recent crisis due to COVID-19.

\subsection{Baseline Algorithms}

We compare our method with a group of baseline algorithms. The first set of baseline models are reallocation strategies adopted by many pension funds. These strategies assign a fixed allocation ratio to relevant assets and rebalance portfolios annually to maintain these ratios. Investors can select a portfolio based on their risk preferences. In general, portfolios weighted more on equities would deliver better performance at the expense of larger volatility. In this work, we consider four such strategies: Allocation 1 (25\% shares, 25\% bonds, 25\% commodities and 25\% volatility index), Allocation 2 (50\% shares, 10\% bonds, 20\% commodities, and 20\% volatility index), Allocation 3 (10\% shares, 50\% bonds, 20\% commodities, and 20\% volatility index), and Allocation 4 (40\% shares, 40\% bonds, 10\% commodities and 10\% volatility index).

The second set of comparison models are mean-variance optimisation (MV) \cite{markowitz1952portfolio} and maximum diversification (MD) \cite{theron2018maximum}. We use moving averages with a rolling window of 50 days to estimate the expected returns and covariance matrix. The portfolio weights are updated at a daily basis and we select weights that maximise Sharpe ratio for MV. The last baseline algorithm is the diversity-weighted portfolio (DWP) from Stochastic Portfolio Theory presented in \cite{samo2016stochastic}. The DWP relates portfolio weights to assets' market capitalisation and it has been suggested to be able to outperform the market index with certainty \cite{fernholz2005diversity}.

\subsection{Training Scheme}

In this work, we use a single layer of LSTM connectivity, with 64 units, to model the portfolio weights and thence to optimise the Sharpe ratio. We purposely keep our network simple to indicate the effectiveness of this end-to-end training pipeline instead of carefully fine-tuning the ``right'' hyperparameters. Our input contains close prices and daily returns for each market index and we take the past 50 days of these observations to form a single input. We are aware that returns can be derived from prices, but keeping returns help with the evaluation of Equation~\ref{eq:return_vol} and we can also treat them as momentum features in \cite{moskowitz2012time}. As our focus is not on feature selection, we choose these commonly used features in our work. The Adam optimiser \cite{kingma2014adam} is used for training our network, and the mini-batch size is 64. We take 10\% of any training data as a separate validation-set to optimise hyperparameters and control overfitting problems. Any hyperparameter optimisation is done on the validation set, leaving the test data for the final performance evaluation and ensuring the validity of our results. In general, our training process stops after 100 epochs. 

\subsection{Experimental Results}

When reporting the test performance, we include transaction costs and use volatility scaling \cite{moskowitz2012time, lim2019enhancing, zhang2020deep} to scale our positions based on market volatility. We can set our own volatility target and meet expectations of investors with different risk preferences. Once volatilities are adjusted, our investment performances are mainly driven by strategies instead of being heavily affected by markets. The modified portfolio return can be defined as:
\begin{equation}
\label{eq:return_vol}
R_{p,t} =  \sum_{i}^n  \frac{\sigma_{tgt}}{\sigma_{i, t-1}}  w_{i,t-1} \cdot r_{i,t} - C \cdot \sum_{i}^n \Big | \frac{\sigma_{tgt}}{\sigma_{i, t-1}}w_{i,t-1} - \frac{\sigma_{tgt}}{\sigma_{i, t-2}}w_{i,t-2} \Big | 
\end{equation}
where $\sigma_{tgt}$ is the volatility target and $\sigma_{i, t-1}$ is an ex-ante volatility estimate of asset $i$ calculated using an
exponentially weighted moving standard deviation with a 50-day window on $r_{i,t}$. We use daily changes of traded value of an asset to represent transaction costs, which is calculated by the second term in Equation~\ref{eq:return_vol}. $C$ (=1bs=0.0001) is the cost rate and we change it to reflect how our model performs under different transaction costs. 

To evaluate the performance of our methods, we utilise following metrics: expected return ($E(R)$), standard deviation of return ($\mathrm{Std}(R)$), Sharpe ratio \cite{sharpe1994sharpe}, downside deviation of return ($\mathrm{DD}(R)$) \cite{mcneil2015quantitative}, and Sortino ratio \cite{sortino1994performance}. All of these metrics are annualised, and we also report on maximum drawdown (MDD) \cite{chekhlov2005drawdown}, percentage of positive return (\% of + Ret) and the ratio between positive and negative return (Ave. P / Ave. L).

Table~\ref{tb:metrics} presents the results of our model (DLS) compard to other baseline algorithms. The top of the table shows the results without using volatility scaling, and we can see that our model (DLS) achieves the best Sharpe's ratio and Sortino ratio, delivering the highest return per risk. However, given the large differences in volatilities, we can not directly compare expected and cumulative returns for different methods, thereby volatility scaling also helps to make fair comparisons.  

\begin{figure}[b]
\begin{subfigure}{0.33\textwidth}
\includegraphics[width=\linewidth, height=3.6cm]{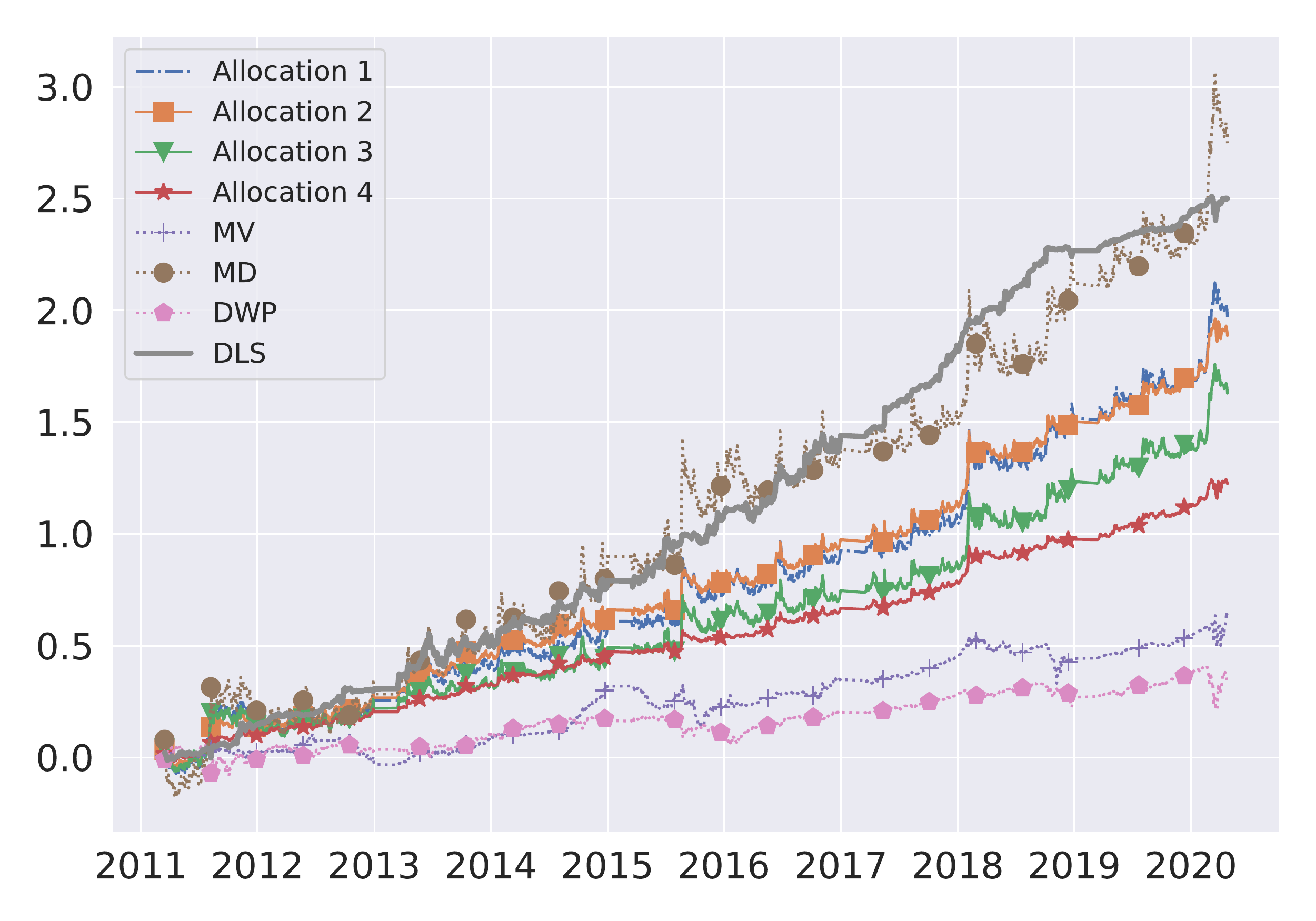} 
\end{subfigure}
\begin{subfigure}{0.33\textwidth}
\includegraphics[width=\linewidth, height=3.6cm]{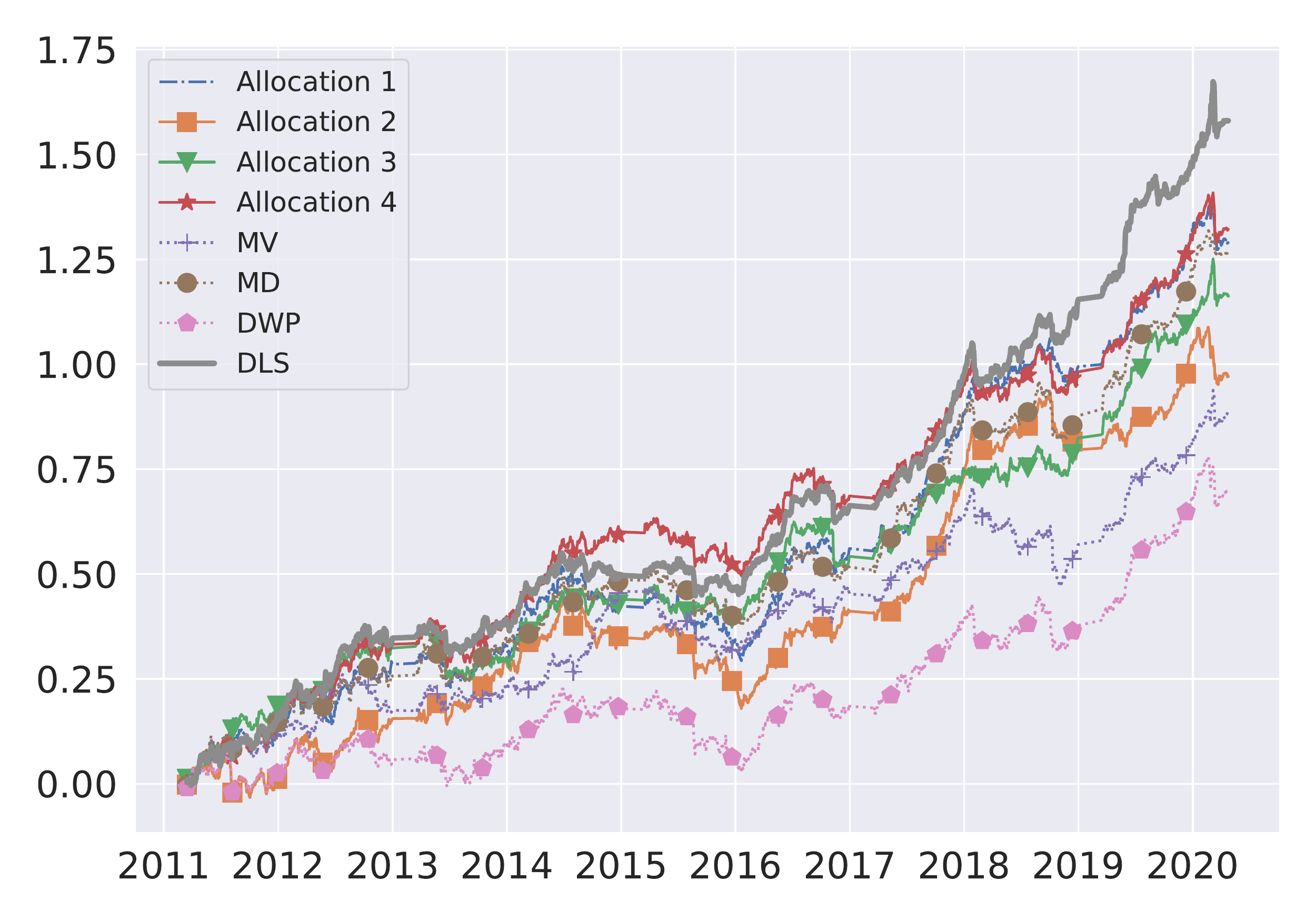}
\end{subfigure}
\begin{subfigure}{0.33\textwidth}
\includegraphics[width=\linewidth, height=3.6cm]{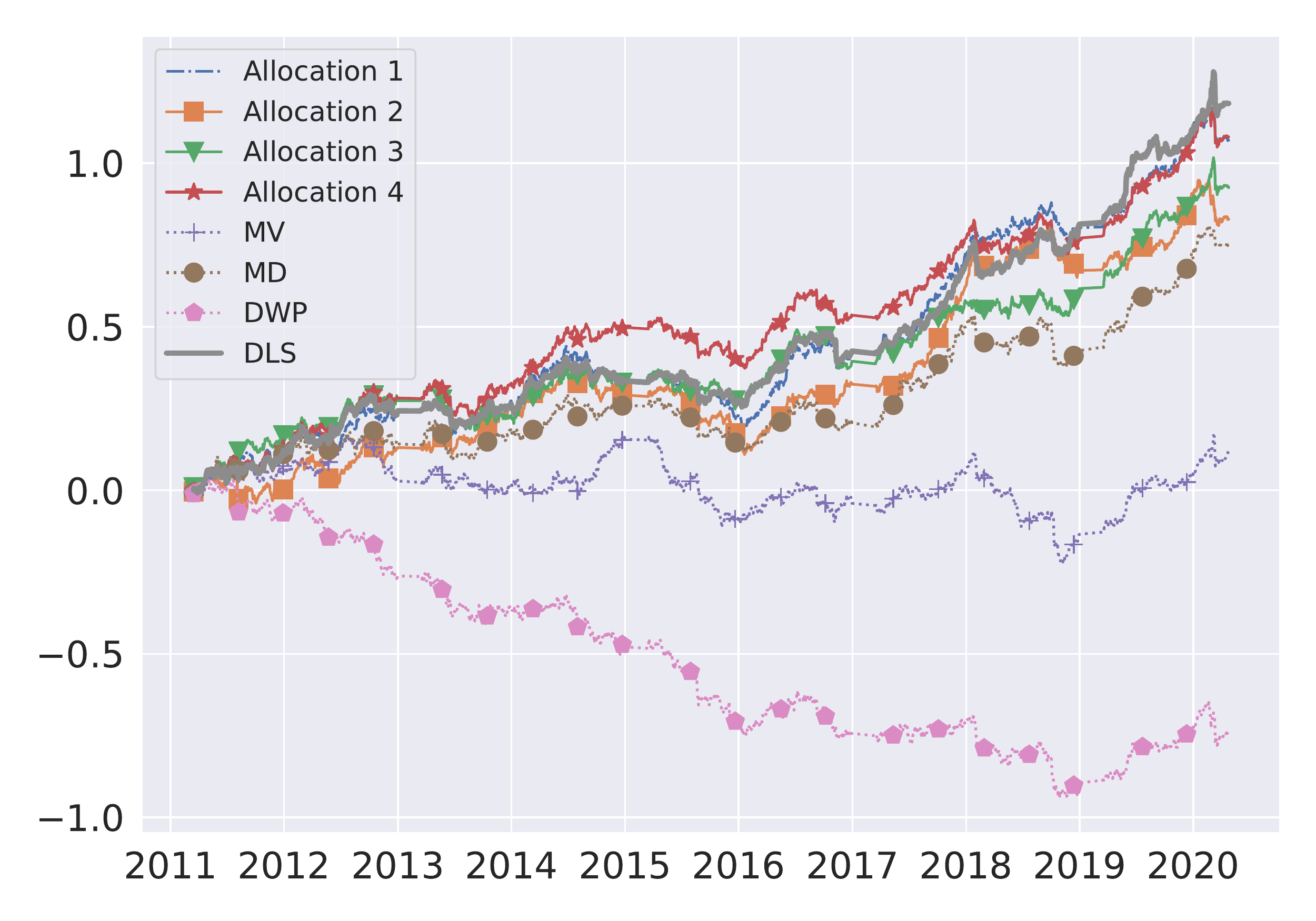}
\end{subfigure}
\caption{Cumulative returns (logarithmic scale) for \textbf{Left}: no volatility scaling and $C$ = 0.01\%; \textbf{Middle}: volatility scaling ($\sigma_{tgt}=0.10$) and $C$ = 0.01\%; \textbf{Right}: volatility scaling ($\sigma_{tgt}=0.10$) and $C$ = 0.1\%.}
\label{fig:cum_return}
\end{figure}

Once volatilities are scaled (shown in the middle of Table~\ref{tb:metrics}), DLS delivers the best performance across all evaluation metrics except for a slightly larger drawdown. If we look at the cumulative returns in Figure~\ref{fig:cum_return}, DLS exhibits outstanding performance over the long haul and the maximum drawdown is reasonable, ensuring the confidence of investors to hold through hard times. Further, if we look at the bottom of Table~\ref{tb:metrics} where a large cost rate ($C$ = 0.1\%) is used, our model (DLS) stills delivers the best expected return and achieves the highest Sharpe and Sortino ratios. 

\begin{table}[h]
\caption{Experiment results for different algorithms.}
\begin{tabular}{l|llllllll}
\toprule
                     & \textbf{E(R)}   &  \textbf{Std(R)} & \textbf{Sharpe} & \textbf{DD(R)}  & \textbf{Sortino} & \textbf{MDD}   & \textbf{\% of + Ret} & \textbf{$\frac{\text{Ave. P}}{\text{Ave. L}}$}\\
\midrule                     
\multicolumn{9}{c}{\textbf{No volatility scaling and $C$ = 0.01\%} }        \\
\midrule
Allocation 1    & 0.282  & 0.303  & 0.929  & 0.136    & 2.065   & 0.142 & 0.479   & 1.193     \\
Allocation 2    & 0.249  & 0.212  & 1.173  & 0.095    & 2.616   & 0.097 & 0.483   & 1.254  \\
Allocation 3    & 0.228  & 0.256  &0.890  & 0.116    & 1.962   & 0.122 & 0.476       & 1.183   \\
Allocation 4    & 0.152  & 0.123  & 1.228  & 0.052    &2.932   & \textbf{0.081} &0.505 & 1.349  \\
MV     & 0.082  & 0.108  & 0.759  &0.069    & 1.192  &0.195 & \textbf{0.562}       & 1.199         \\
MD     &0.462  & 0.523  & 0.882  & 0.239    & 1.931  & 0.273 & 0.473       & 1.182         \\
DWP & 0.051  & \textbf{0.102}  & 0.493  & \textbf{0.067}    & 0.740   & 0.179 & 0.549 & 1.107        \\
DLS   & \textbf{0.313}  & 0.168  & \textbf{1.858} & 0.099    & \textbf{3.135}  &0.102 & 0.537    & \textbf{1.518}  \\
\midrule
\multicolumn{9}{c}{\textbf{Volatility scaling ($\sigma_{tgt}=0.10$) and $C$ = 0.01\%}}               \\
\midrule
Allocation 1    & 0.160  & 0.105  &1.526  & 0.061    & 2.629  & 0.111 & 0.554    & 1.289      \\
Allocation 2    & 0.123  & 0.106  &1.146  & 0.065    & 1.861   &0.127& 0.549      & 1.211     \\
Allocation 3    &0.145  & 0.105  & 1.383  & 0.061    & 2.396   & \textbf{0.105} &0.542   & 1.259     \\
Allocation 4   & 0.164  & 0.104  & 1.579  & 0.064   & 2.588   & 0.112 & 0.565    & 1.303     \\
MV        & 0.112  & 0.100  & 1.120  & 0.063    & 1.767  & 0.211 & 0.561       & 1.213        \\
MD     &0.157  & 0.106 & 1.484  &0.065    & 2.414   & 0.125 & 0.565       &1.297        \\
DWP & 0.089  & 0.109  & 0.818 & 0.069   & 1.291  & 0.115 & 0.556       & 1.148         \\
DLS   & \textbf{0.206}  &0.105  & \textbf{1.962}  & 0.062    & \textbf{3.322}   & 0.123 & \textbf{0.559}       & \textbf{1.375}         \\
\midrule
\multicolumn{9}{c}{\textbf{Volatility scaling ($\sigma_{tgt}=0.10$) and  $C$ = 0.1\%}}                \\
\midrule
Allocation 1    & 0.133  & 0.105  & 1.274 & 0.061    & 2.172   & 0.113& 0.548       & 1.236         \\
Allocation 2    & 0.105  &0.107 & 0.986  & 0.066    & 1.590   & 0.244 & 0.547       & 1.179         \\
Allocation 3    & 0.117  & 0.105  & 1.110 & 0.061   & 1.903   &\textbf{0.107} & 0.538       & 1.203         \\
Allocation 4    & 0.135  & 0.104  & 1.299  & 0.064    & 2.108  & 0.114 & \textbf{0.559}       & 1.244         \\
MV        &0.019  & 0.101  & 0.191  & 0.066    &0.293   & 0.324 & 0.537       & 1.033       \\
MD     & 0.095  & 0.106 & 0.899  & 0.066    & 1.431   & 0.145 & 0.549       & 1.171         \\
DWP & -0.083 & 0.110  & -0.753 & 0.074    & -1.129  & 0.627 & 0.508       & 0.880         \\
DLS           & \textbf{0.148} & 0.105  & \textbf{1.403}  & 0.063    & \textbf{2.327}   & 0.125 & 0.547       & \textbf{1.272}        \\
\bottomrule
\end{tabular}
\label{tb:metrics}
\end{table}

\begin{figure}[t]
\centering
\includegraphics[width=5in, height=2in]{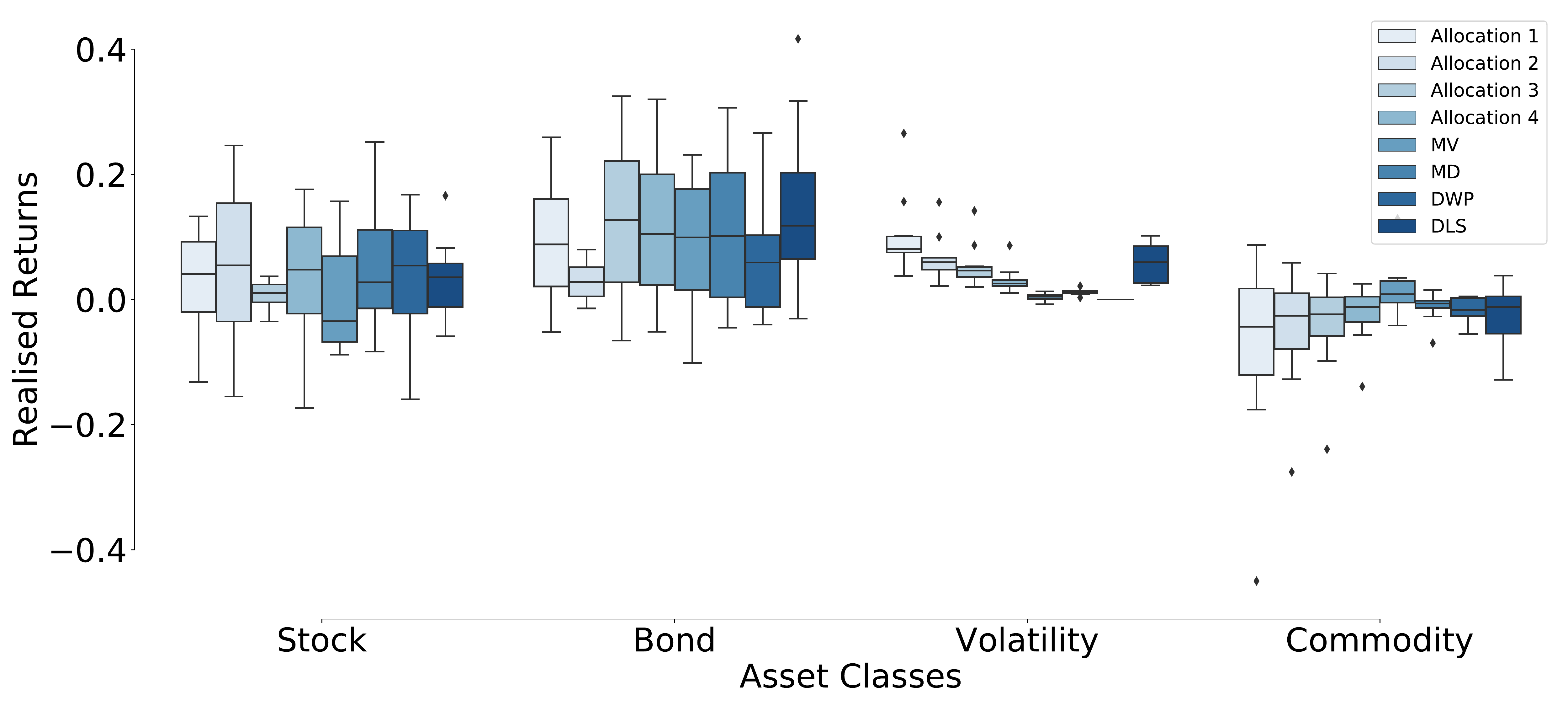}
\includegraphics[width=5in, height=2in]{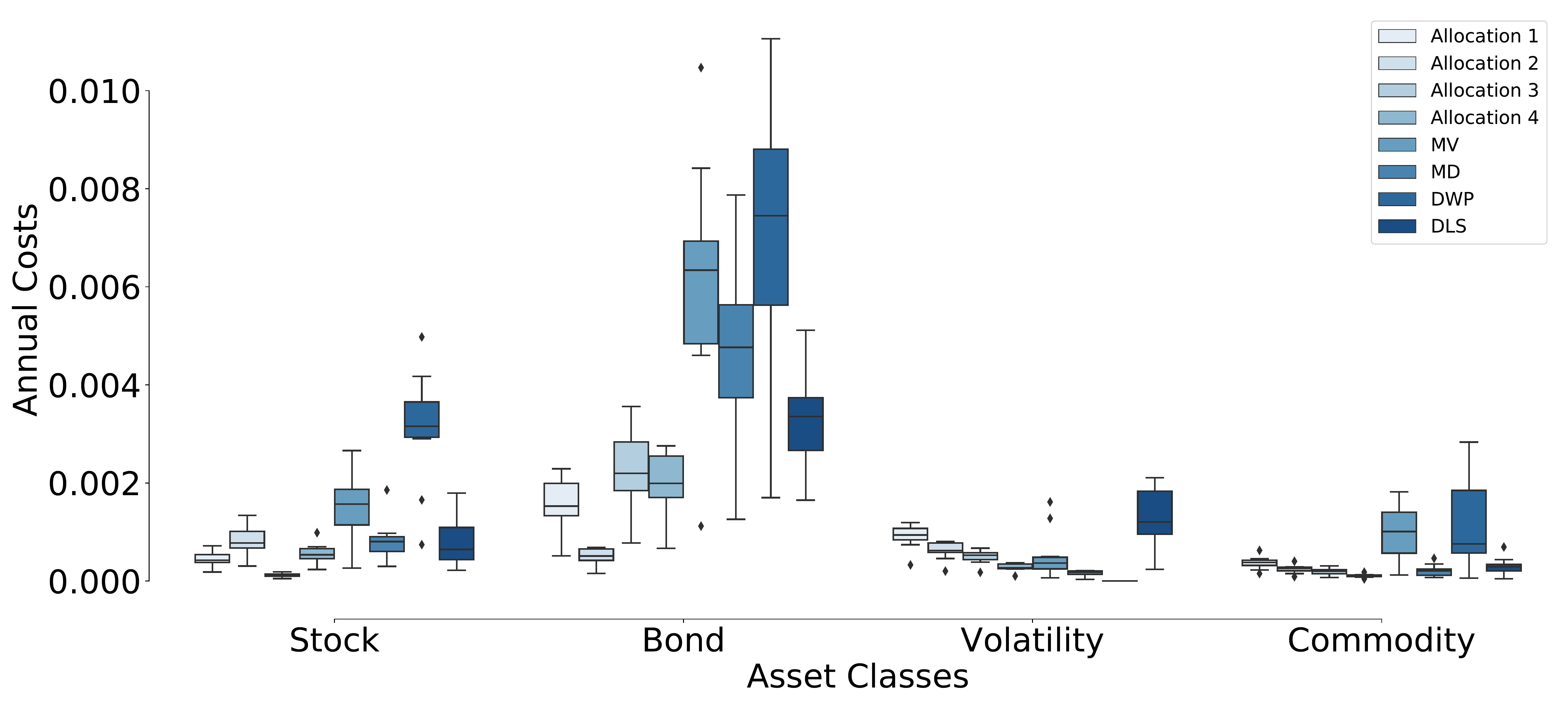}
\caption{Boxplot for \textbf{Top}: annual realised trade returns; \textbf{Bottom}: annual accumulated costs for different assets with volatility scaling ($\sigma_{tgt}=0.10$) and $C$ = 0.01\%.}
\label{fig:boxplot}
\end{figure}

However, with a higher cost rate, we can see that reallocation strategies work well and, in particular, Allocations 3 and 4 achieve comparable results to our method. In order to investigate why performance gap diminishes with a higher cost rate, we present the boxplots for annual realised trade returns and accumulated costs for different assets in Figure~\ref{fig:boxplot}. Overall, our model delivers better realised returns than reallocation strategies, but we also accumulate much larger transaction costs since our positions are adjusted on a daily basis, leading to a higher turnover. 

For reallocation strategies, daily position changes are only updated for volatility scaling. Otherwise, we only actively change positions once a year to rebalance and maintain the allocation ratio. As a result, reallocation strategies deliver minimal transaction costs. This analysis aims to indicate the validity of our results and show that our method can work under unfavorable conditions.

\subsection{Model Performance during 2020 Crisis }

Due to the recent COVID-19 pandemic, global stock markets fell dramatically and experienced extreme volatility. The crash started on the 24th February 2020 where markets reported their largest one-week declines since the 2008 financial crisis. Later on, with an oil price war between Russia and the OPEC countries, markets further dampened and encountered the largest single-day percentage drop since Black Monday in 1987. As of March 2020, we have seen a downturn of at least 25\% in the US markets and 30\% in most G20 countries. The crisis shattered many investors' confidence and resulted in a great loss of their wealths. However, it also provides us with a great opportunity to stress test our method and understand how our model performs during the crisis.  

\begin{figure}[h]
\centering
\includegraphics[width=5.5in, height=3.5in]{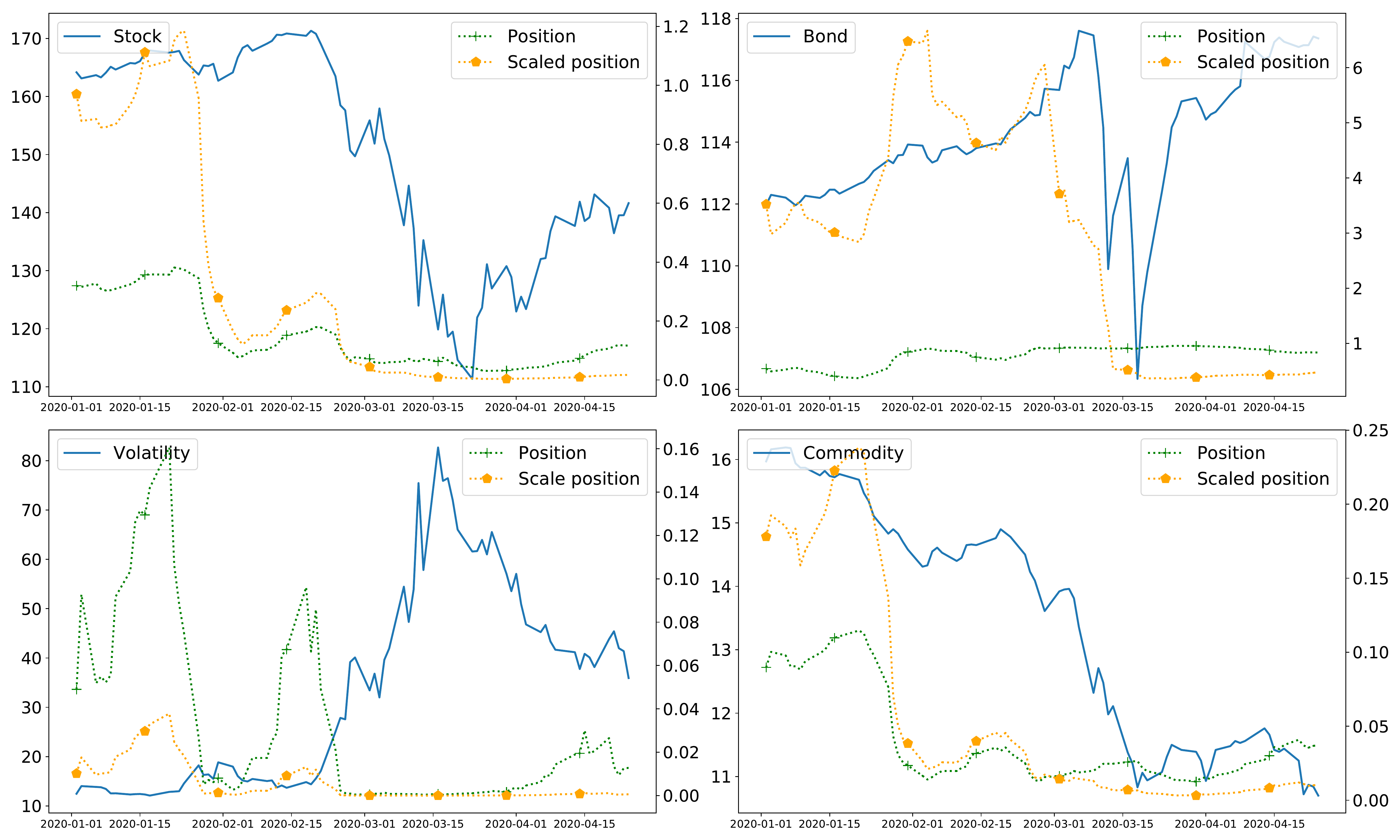}
\caption{Shifts of portfolio weights for our model (DLS) during the crisis of COVID-19 with volatility scaling ($\sigma_{tgt}=0.10$).}
\label{fig:action}
\end{figure}

In order to study the model behaviours, we plot how our algorithm allocated the assets from January to April 2020 in Figure~\ref{fig:action}. At the beginning of 2020, we can see that our model had a quite diverse holding. However, after a small dip in stock index in early February, we almost had only bonds in our portfolio. There were some equity positions left but very small positions for volatility and commodity indices. When the crash started on 24th February, our holdings were concentrated on the bond index which is considered to be safe assets during the crisis. Interestingly, the bond index also fell this time (in the middle of March) although it rebounded quite quickly. During the bond falling, our original positions did not change much but the scaled positions decreased a lot for the bond index due to a spiking volatility, therefore our drawdown was small. Overall, we can see that our model delivers reasonable allocations during the crisis and our positions are protected through volatility scaling.


\subsection{Sensitivity Analysis}

\begin{figure}[ht]
\centering
\includegraphics[width=5.5in, height=2.5in]{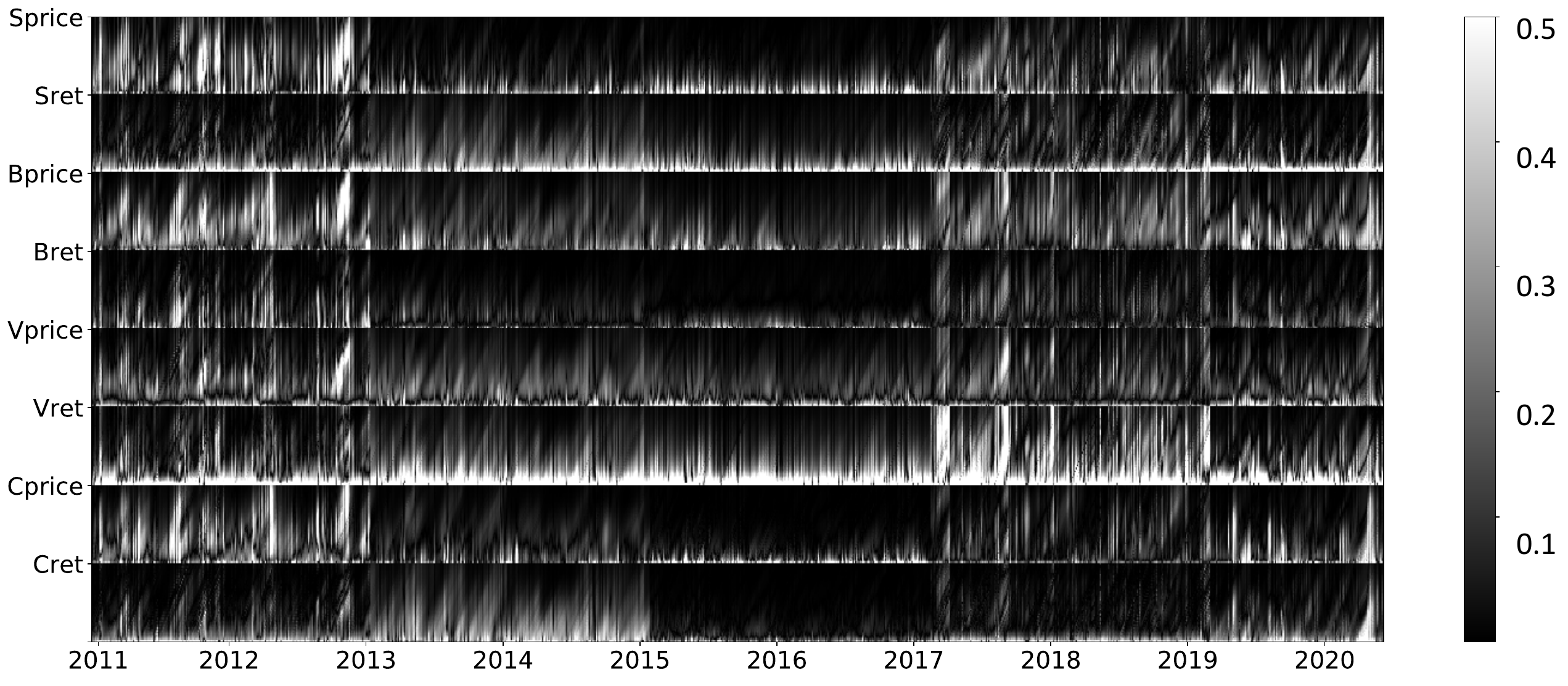}
\caption{Sensitivity analysis for input features over the time.}
\label{fig:sensitivity}
\end{figure}

In order to understand how input features affect our decisions, we 
study the sensitivity analysis presented in \cite{moody2001learning} for our method. The absolute normalised sensitivity of feature $x_i$ is defined as:
\begin{equation}
S_i =\frac{\frac{dL}{dx_i}}{\text{max}_j \left | \frac{dL}{dx_j} \right |}
\end{equation}
where $L$ represents the objective function and $S_i$ captures the relative sensitivity for feature $x_i$ compared with other features. We plot the time-varying sensitivities for all features in Figure~\ref{fig:sensitivity}. The y-axis indicates the 400 features we have because we use 4 indices (each with prices and returns) and we take a timeframe of past 50 observations to form a single input so there are 400 features in total. The row labeled ``Sprice'' represents price features for the stock index and the bottom of row ``Sprice'' means the most recent price for that observation. Same convention is used for all other features. 

The importance of features varies over the time, but the most recent features always make the biggest contributions as we can see that the bottom of each feature row has the highest weight. This observation meets our understanding as, for time-series, recent observations carry more information. The further away from the current observation point, the less importance of features show and we could adjust features used based on this observation such as using a small lookback window.

\section{Conclusion}
\label{conclusion}

In this work, we adopt deep learning models to directly optimise a portfolio's Sharpe ratio. This pipeline bypasses the traditional forecasting step and allows us to optimise portfolio weights by updating model parameters through gradient ascent. Instead of using individual assets, we focus on ETFs of market indices to form a portfolio. Doing this substantially reduces the scope of possible assets to choose from, and these indices have shown robust correlations. In this work, four market indices have been used to form a portfolio. 

We compare our method with a wide range of popular algorithms including reallocation strategies, classical mean-variance optimisation, maximum diversification and stochastic portfolio theory model. Our testing period is from 2011 to the April of 2020, and include the recent crisis due to COVID-19. The results show that our model delivers the best performance and a detailed study of our model performance during the crisis shows the rationality and practicability of our method. A sensitivity analysis is included to understand how input features contribute to outputs and the observations meet our econometric understanding, showing the most recent features are most relevant.  

In subsequent continuation of this work, we aim to study portfolios performance under different objective functions. Given the flexible framework of our approach, we can maximise Sortino ratio or even the diversification degree of a portfolio as long as functions are differentiable. We further note that the volatility estimates used for scaling are lagged estimates that do not necessarily represent current market volatilities. We consider another extension to this work to thus adapt the network architecture to infer (future) volatility estimates as a part of the training process. 

\section*{Acknowledgements}

The authors would like to thank members of Machine Learning Research Group at the University of Oxford for their useful comments. We are most grateful to the Oxford-Man Institute of Quantitative Finance for support and data access.


\end{document}